# Reducing cross-flow vibrations of underflow gates: experiments and numerical studies

C.D. Erdbrink
University of Amsterdam, The Netherlands
Deltares, The Netherlands
National Research University of Information Technologies, Mechanics and Optics, Saint Petersburg, Russia

V.V. Krzhizhanovskaya
National Research University of Information Technologies, Mechanics and Optics, Saint Petersburg, Russia
Universtiy of Amsterdam, The Netherlands

P.M.A. Sloot
Universtiy of Amsterdam, The Netherlands
National Research University of Information Technologies, Mechanics and Optics, Saint Petersburg, Russia
Nanyang Technological University, Singapore

*Abstract*
An experimental study is combined with numerical modelling to investigate new ways to reduce cross-flow vibrations of hydraulic gates with underflow. A rectangular gate section placed in a flume was given freedom to vibrate in the vertical direction. Holes in the gate bottom enabled leakage flow through the gate to enter the area directly under the gate which is known to play a key role in most excitation mechanisms.
For submerged discharge conditions with small gate openings the vertical dynamic support force was measured in the reduced velocity range $1.5 < Vr < 10.5$ for a gate with and without holes. The leakage flow through the holes significantly reduced vibrations. This attenuation was most profound in the high stiffness region at $2 < Vr < 3.5$.
Two-dimensional numerical simulations were performed with the Finite Element Method to assess local velocities and pressures for both gate types. A moving mesh covering both solid and fluid domain allowed free gate movement and two-way fluid-structure interactions. Modelling assumptions and observed numerical effects are discussed and quantified. The simulated added mass in still water is shown to be close to experimental values. The spring stiffness and mass factor were varied to achieve similar response frequencies at the same dry natural frequencies as in the experiment. Although it was not possible to reproduce the vibrations dominated by impinging leading edge vortices (ILEV) at relatively low $Vr$, the simulations at high $Vr$ showed strong vibrations with movement-induced excitation (MIE). For the latter case, the simulated response reduction of the ventilated gate agrees with the experimental results. The numerical modelling results suggest that the leakage flow diminishes the whipping effect of fluctuations at the trailing edge associated with the streamwise pressure drop across the gate and the body's vertical oscillatory motion.

## 1. Introduction

This study presents a novel hydraulic gate design aimed at reducing vibrations induced by underflow. The dynamic response of a hydraulic gate due to its interaction with the flow strongly depends on details of the gate bottom geometry. Numerous experimental studies of flow-induced vibrations (FIV) of gates have previously looked into the characteristics of gate shapes (Hardwick 1974, Vrijer 1979, Kolkman 1984). The gained insight in excitation mechanisms has resulted in widespread rules of thumb for unfavourable designs that should be avoided as well as favourable design features (e.g. Thang 1990, Naudascher & Rockwell 1994). However, fundamental knowledge and practical experience have not culminated in one ideal universal shape – partly because the surrounding structure is an important factor. Consequently, hydraulic engineers still stumble on the problem of gate vibrations when designing a new structure or when conditions of gate operation change over time.

Experimental and numerical models are incapable of capturing all degrees of freedom (d.o.f.) experienced by real-life gates (mass-vibration mode in cross-flow and in-flow direction, bending, torsion). Streamwise (horizontal) vibrations are usually studied separately (e.g. Jongeling 1988) and sometimes in combination with the cross-flow mode (Billeter & Staubli 2000). In this study we consider the most frequently investigated mode for a vertical-lift underflow gate: one d.o.f. in the cross-flow direction.

The discharge past a partly lifted gate is driven by a head difference, which causes a streamwise pressure gradient. At sufficiently high downstream water levels, the discharge is submerged and a quasi-stationary rotational cell with horizontal axis exists in the downstream region. The flow accelerates as it approaches the gate; the mean velocity reaches a maximum just past the gate in the point of maximal flow contraction called the vena contracta. Depending on the head difference, submergence and the gate's geometry and position, it experiences a steady positive or negative lift force (Naudascher 1991). This quasi-steady description suffices as long as the gate does not oscillate.

The emergence and severity of flow-related dynamic forces on the gate are related to flow instabilities and body motion effects. To describe and explain relations between flow properties, forces and gate motion during oscillation, several excitation mechanisms were introduced. Periodic fluctuations of the separated flow's shear layer may cause an active response. For gates with a sharp upstream edge, this mechanism is called Impinging Leading Edge Vibrations (ILEV). If the gate bottom has an extending lip in streamwise direction, the shear layer separated from the upstream edge may reattach to the gate bottom in an unstable way, giving dynamic excitation. In a different mechanism, periodic forces are the result of initially small gate movements. This self-exciting process is called Movement-Induced Excitation (MIE). The galloping phenomenon falls in this category.

Previous investigations have proved that most severe vibrations of underflow gates in submerged flow occur at small gate openings and are predominantly caused by ILEV and MIE mechanisms (Hardwick 1974, Thang & Naudascher 1986a and 1986b). The current study therefore focuses on small gate openings and does not look at other mechanisms such as noise excitation. Other notable studies are Kapur & Reynolds (1967), Naudascher & Rockwell (1980), Thang (1990), Kanne et al. (1991), Ishii (1992) and Gomes et al. (2001). Overviews of flow-induced vibrations of gates are found in Kolkman (1976), and Naudascher & Rockwell (1994) and Jongeling & Erdbrink (2009); Blevins (1990) treats FIV in a wider context. None of these authors considered holes in the gate as a means to weaken vibrations.

Assuming that adding structural damping or avoiding critical gate openings are unfeasible options, the shape of the gate bottom is the decisive factor determining the tendency to vibrate. If the flow passes the gate while remaining attached, or if there is a fixed separation point and a stable reattachment, or if the shear layer is kept away from the bottom in all circumstances, then the ILEV mechanism may be avoided. A thin, sharp-edged geometry with separation from the trailing edge is favourable, since potential shear layer instabilities occur downstream from the gate and a small bottom area inhibits the occurrence of large (suction) forces on the gate, thus minimizing the risk of MIE vibrations. But such a design is often not practical or is influenced by the addition of unfavourable details such as flexible rubber seals. The investigation at hand takes the unfavourable thick flat-bottom rectangular gate as a reference gate and introduces leakage flow through a hole in the bottom section as a potential way to improve its vibration properties, see Figure 1.

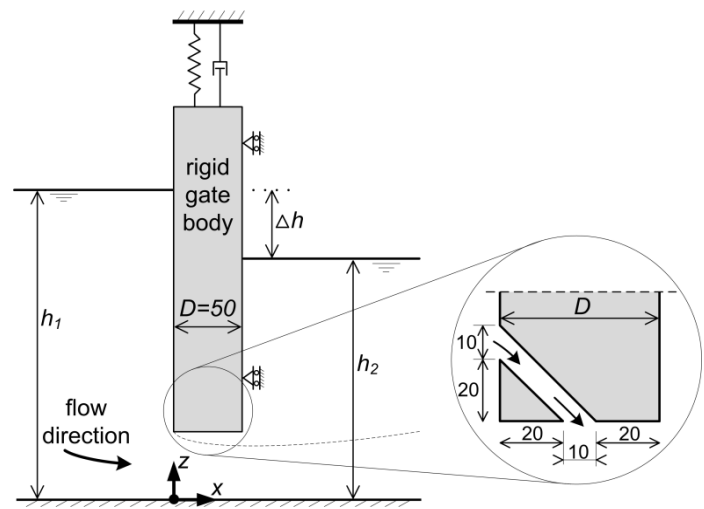

*Figure 1. Sketch of gate configuration and cross-section of ventilated gate design (detail on the right). Measurements in millimeters; not drawn to scale.*

Numerical simulations of gate vibrations based on elementary physics equations are inevitably complex and computationally involved. The computational Fluid-Structure Interaction (FSI) model needs to deal with very small displacements of varying frequencies and should represent the boundary forces accurately. The fact that the flow is incompressible, turbulent and has a free-surface adds to the numerical challenge from a Computational Fluid Dynamics (CFD) viewpoint. Many modelling simplifications (such as lowering the fluid's viscosity or including only partial interaction between fluid and solid) are not permitted as they change the physics of the problem too much so that the excitation mechanisms have no chance to occur. Meshes that do not match up at the fluid-solid interface have the disadvantage of requiring interpolation functions (De Boer 2008) and dedicated load transfer schemes (Jaiman et al 2006). Applying the Finite Element Method (FEM) gives the possibility to contain both the solid and fluid domain in one and the same computational moving mesh – thus avoiding a number of coupling issues. Rugonyi & Bathe (2001) explore FEM for FSI problems with incompressible fluid flow.

Numerical studies of gate dynamics are scarce compared to flow-induced vibrations of bluff bodies. The field of Vortex-Induced Vibrations (VIV), see Sarpkaya (2004), tackles a fundamentally different problem and has a much wider range of applications. The flow around stationary and vibrating cylinders has become a classical numerical modelling problem (Al-Jamal & Dalton 2004, Dai et al. 2013). Numerical modelling of gates often has the goal of verifying the design of a specific hydraulic structure. Determining static support forces is hence of greater concern than addressing vibrations (Scheffermann & Stockstill 2009, Liu et al. 2011). A study on gate control with numerical modelling on different scales, including the influence of the free surface is provided by Erdbrink et al. (2013). Those studies that do consider vibrations more often than not fail to make links to the existing state of the art, thus not optimally contributing to fundamental knowledge (e.g. Lupuleac et al. 2007).

The two aims of this paper are to provide an investigation into the dynamic behaviour of a perforated gate *and* to assess the feasibility of numerical physics-based modelling as a complementary tool next to physical modelling for evaluating flow-induced vibrations of gates. The motivation behind the second aim is the engineer's desire to develop quick assessment tools for gate designs. Moreover, the growing application of early warning systems to flood defense systems (Erdbrink et al. 2012) calls for numerical models capable of foreseeing risky future situations.

The next section describes the set-up and results of the physical scale model experiment. It includes a paragraph on analysis giving definitions of physical parameters. In section 3, the numerical model is treated, showing results of validation runs and simulations of a vibration case from the experiment. Section 4 discusses the excitation mechanisms and the working of the pierced gate. The limitations of the numerical model are discussed as well. The final section contains the conclusions and possible future work.

## *2. Physical experiment*

*2.1 Experimental set-up*
The experiment of a gate section was performed in a 1 m wide and about 90 m long laboratory flume. Appendix A gives sketches of the set-up and the online material contains photographs of the set-up. A straight, vertically placed underflow gate is suspended in a steel frame that is fixed to the flume. The dimensions of the gate are 1100x600x50 mm (height x width x thickness); it is a stiff plate and thus acts as a linear mass-spring oscillator body with one degree of freedom in the cross-flow vertical direction. To prevent measurement equipment and frame parts from influencing the flow, the flume was locally narrowed to 0.5 m by constructing concrete plywood walls and a sloped ramp around the gate. The design ensures that the flow directly upstream from the gate is attached to the walls and has low turbulence intensity.

Two coil springs with low stiffness carry most of the static loads in vertical direction. The dynamic loads are mostly carried by one stiff central spring with an adjustable stiffness to enable controlled variation of the natural frequency. For the two weak side springs, linear coil springs (Alcomex TR-1540) were used;

for the stiff main spring a double leaf spring was custom built with high yield strength steel and a high elastic limit (Armco 17-7PH, hardening condition TH1050).

Five horizontal supports, three hinged steel rods in longitudinal direction and two in cross direction, enable vertical movement of the gate while fixing the gate position horizontally. Six force meters were installed: vertically one for each spring and horizontally one for each longitudinal support. Only the main vertical support force is used in the analysis of the gate response. The sample frequency was 200 Hz. The length of the analysed data files was 90 s on average and had a minimum of 60 s, yielding a frequency resolution of at least 0.0017 Hz. Recording started after reaching equilibrium water levels and horizontal support forces. The signal analysis done in MATLAB consisted of standard FFT-analysis using sliding windows. Furthermore, the discharge and the water levels on the upstream and the downstream side of the gate were measured using resistance-type water level meters. Erdbrink (2012a) gives more details on the experiment and analysis method.

As shown in Figure 1, two gate types were tested: the flat rectangular-shaped bottom (with smooth surface, sharp edges and without extending lip) will be called 'original' or 'closed' gate. The adapted gate with holes pierced though the bottom section will be called 'new' or 'ventilated gate'. There was a slight difference in mass between the two gates due to removal of material to create the holes and a hollow space inside the bottom element that filled with water after making the holes. The total masses in wet condition were 17.2 kg for the closed gate and 17.3 kg for the ventilated gate.

The two bending blades of the main spring have dimensions 600 mm x 30 mm x 4 mm ($l$ x $b$ x $t$). The bending length $L$ (< 600 mm) can be adjusted by movable clamps, thus varying structural stiffness. See Appendix A. Linearity of this spring was confirmed by static loading tests for different bending lengths. The main spring is installed in parallel with two side springs, each with constant $k_{side}$ = 0.57 N/mm. The relation between bending length and total stiffness of the three springs is derived from constitutive relations plus Hookes law:

$$k_{total} = k_{main} + 2k_{side} = \frac{8Ebt^3}{L^3} + 1.14$$

with $k_{total}$ the combined spring stiffness in N/mm, $E$ the modulus of elasticity in N/mm², $b$ the width and $t$ the thickness of the leaf spring blades in mm and $L$ the length between the clamps in mm. This formula was calibrated with free vibration tests in air to increase accuracy between chosen $L$ and achieved $k_{total}$ and dry natural frequency $f_0$. During the experiment the stiffness was varied between and 19.3 N/mm and 967 N/mm, corresponding to a range in achieved undamped natural frequency in air of 5.33 Hz < $f_0$ < 37.7 Hz for the original gate with closed bottom section. Damping was not a varied measurement parameter in the experiment. Estimated from free vibration tests in still water, the damping ratio $\gamma = c/2\sqrt{km}$ was between 0.035 and 0.19. Comparison with vibration in air at equal stiffness indicates that hydraulic damping is dominant over structural damping (at moderate amplitudes). Hence, the Scruton number $S_c = 2\gamma m/\rho_w D^2$ was in the range 0.48 < $S_c$ < 2.6.

*2.2 Analysis and definitions*

The motion equation in vertical $z$-direction for partly submerged bodies has to include hydrodynamic or 'added' coefficients (see e.g. Kolkman 1980):

$$(m + m_w)\ddot{z} + (c + c_w)\dot{z} + (k + k_w)z = F(t, z, \dot{z})$$

Here, $m$ is mass, $c$ is damping, $k$ is stiffness, $F$ is the excitation force and the subscript $w$ indicates the added coefficients. The dependency of $F$ on displacement and velocity represent non-linear coupling. For the same body vibrating in air, we have (Den Hartog 1956)

$$\ddot{z} + 2\gamma\omega_0\dot{z} + \omega_0^2 z = F/m \, ,$$

using the damping ratio $\gamma$ and the undamped natural angular frequency in air $\omega_0 = \sqrt{k/m} = 2\pi f_0$. For the damped case, $\omega_0$ and $f_0$ have to be multiplied by a factor $\sqrt{1-\gamma^2}$. The excitation is a time-dependent hydraulic force which can be written as

$$F = \frac{1}{2} C_F \rho U^2 W D = C_F F_0$$

With $W$ the cross-flow width of the gate section on which $F$ works, $F_0$ a stationary reference fluid force and $C_F$ a force coefficient equal to

$$C_F = C_F' \sin(\omega t + \varphi)$$

Where $\varphi$ is the phase shift between excitation and displacement. Note that in this study not the displacement $z$ but the response force $F_z$ is measured, the amplitude of which is denoted $\widehat{F}_z$. Furthermore, for the pressure on the gate bottom boundary $p_{bound}$ we use the pressure coefficient $C_P$ defined logically as $p_{bound}$ divided by $\rho g \Delta h$. The two-dimensional discharge formula for an underflow gate section in submerged flow is

$$q = C_D a U = C_D a \sqrt{2g\Delta h}$$

With $q$ discharge per unit width in m³/s/m or m²/s, $C_D$ the dimensionless discharge coefficient for submerged flow, $a$ the lifting height or gate opening, $U$ the flow velocity in the vena contracta, estimated with Bernoulli's formula, in m/s and $\Delta h = h_1 - h_3$. The reduced velocity $V_r$ is used as dimensionless descriptive quantity for the flow-induced vibrations. It is defined here as

$$V_r = \frac{\sqrt{2g\Delta h}}{f_z D}$$

Where $f_z$ is the dominant response frequency, the numerator represents the characteristic flow velocity and the gate thickness in flow direction $D$ (see Figure 1) is taken as characteristic length scale. In the present study, the added mass $m_w$ is estimated experimentally in still water by free vibration tests in air and still water. It follows from

$$\frac{f_{0,water}}{f_0} = \sqrt{\frac{1 + k_w/k}{1 + m_w/m}}$$

because the added rigidity (hydraulic stiffness) $k_w$ due to buoyancy on the submerged part of the gate is found via Hooke's law: $k_{w,buoyancy} = \rho g W D$. The error made by neglecting damping here is less than 2%. Numerically, $m_w$ near a wall in still water was computed in a potential flow model with a finite difference method by Kolkman (1988), and studied many times since then in the context of gates (e.g. Anami et al. 2012). A very universal approach is the observation in Kolkman (1984) that the total kinetic energy of the fluid can be expressed as

$$E_{kin} = \frac{1}{2} m_w \dot{z}^2$$

Summing the velocity magnitude over all computational nodes should yield a value for $m_w$, assuming that the object velocity is known.

*2.3 Measurement conditions and variation of parameters*
The present study is focused on the critical range for cross-flow vibrations: 98% of the measurements lie in the interval $0.48 \leq a/D \leq 1.50$. The gate submergence $C_s = (h_2 - a)/D$, with $h_2$ the water depth measured downstream from the gate was in the range $4.2 < C_s < 6.2$, for 98% of the data. This means flow conditions were close to fully submerged with minor free surface fluctuations. Discharge coefficient $C_D$ is estimated from the measured pump discharge to be on average 0.80 with standard deviation 0.11 for the original gate and on average 0.83 with standard deviation 0.10 for the ventilated gate. The achieved $V_r$ ranges were $1.2 < V_r < 11.6$ for the original gate and $1.8 < V_r < 9.5$ for the modified gate. The Reynolds number defined as $Re = UD/\nu$, again with $U = (2g\Delta h)^{0.5}$, was in the range $3.2 \cdot 10^4 < Re < 1.3 \cdot 10^5$.

*2.4 Results of physical experiment*
The focus of the experimental data analysis is on determining how dominant force amplitudes in cross-flow direction change with $V_r$ for both gate types. Absolute maximum force amplitudes depend on structural damping of a particular set-up and are of less interest; therefore response amplitudes are presented for different settings relative to the stationary hydrodynamic force $F_0$.

The plots in Figure 2 show the results: the dimensionless dynamic force response of the closed gate and the gate with holes. Judging from the plot, the response may be divided into three different regions: $2 < V_r < 3$ (relatively high stiffness), $3 < V_r < 8.5$ (medium stiffness) and $V_r > 8.5$ (relatively low stiffness). Response values of $\hat{F}_z > 5N$ represent significant, regular oscillations with response frequencies in the range $4.6$ Hz $< f_{resp,z} < 20.2$ Hz. Vertical displacement amplitudes estimated from the force amplitude as $\hat{z} = \hat{F}_z/k$ were overall less than $0.1D$. The strongest recorded force response amplitude in the high stiffness region was found at $V_r = 2.54$. The maximum response in the relatively low stiffness vibration region occurred at $V_r = 10.16$. The excitation mechanisms associated with these two regions are discussed in section 4.1.

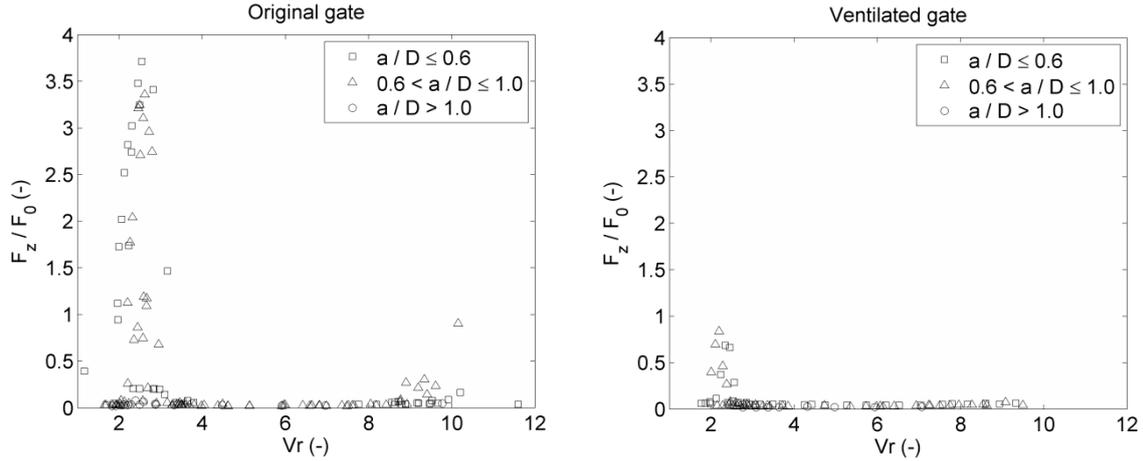

Figure 2. Overview of vibration characteristics of the original rectangular gate (left) and the new ventilated gate (right): reduced velocity $V_r$ versus dimensionless vibration amplitude $\widehat{F}_z/F_0$ of the main vertical force meter; a is the gate opening, gate thickness D is 50 mm.

The results show that the vibrations found at low $V_r$ occur quite suddenly. There are steep increases in force amplitude around $V_r = 2$ and $V_r = 3.0$–$3.5$; most significantly for gate openings less than or equal to $D$. Tests at gate openings smaller than $0.5D$ were hindered by the risk of the gate hitting the flume bottom. Although the gate opening was not varied over a large range, the data seems to show that the force response at $V_r \approx 2.5$ occurs at a smaller gate opening than for the response maximum at $V_r \approx 10$.

Zooming in on the densely sampled area of relatively high frequency response at $2 < V_r < 3$ gives further insights (Figure 3). The measurements were done by making series of about ten data points of different stiffness settings, keeping gate opening and discharge constant.

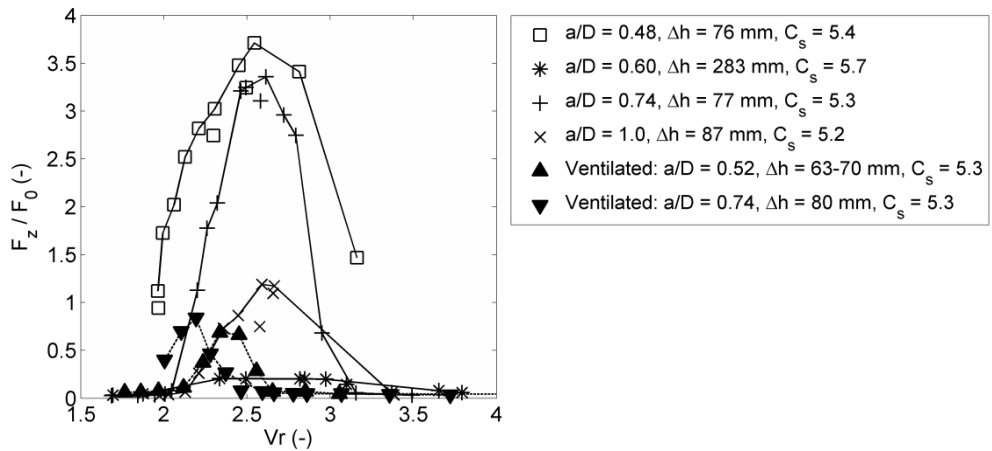

Figure 3. The most significant vibrations at $1.5 < V_r < 4$. Measurements series represent constant gate opening and head difference for varying suspension stiffness. The solid triangles mark the gate

*with holes, all other symbols denote the gate without holes. Responses with $\hat{F}_z/F_0 < 0.25$ not belonging to these series are left out for clarity.*

These results reconfirm that in this reduced velocity region significant cross-flow vibrations occur at gate openings in the range $0.5D \leq a \leq D$. For $1.5 < V_r < 4$, with the strongest force amplitudes around $a/D = 0.5$. From Figure 3 it appears that for head differences $\Delta h \approx 1.5D$–$1.75D$, vibration maxima decrease with increasing gate opening. The series with high head difference $\Delta h \approx 5.7D$ differs from the rest because of its higher velocities and applied stiffness. This most likely resulted in a higher achieved structural damping, whence the lower response.

Despite the smaller data set size of the new gate design, the overall effect of applying ventilation holes seems clear. For very similar conditions at $1.5 < V_r < 3.5$, the force amplitude of the vibration is a factor 3 less for the rectangular gate with added holes. For $3.5 < V_r < 9.5$, the holed gate shows a very low response for similar conditions (openings and head difference) as the original gate. From the available data points at $V_r > 8.5$, the picture of the effect of adding holes is incomplete. More data is needed for $V_r \geq 10$. The available data suggests that the ventilation succeeds in having a mitigating effect on the vibrations in that range as well.

The ventilated gate shows an overall shift towards lower achieved $V_r$-values compared to the original gate. The maximum force response of the new gate lies at $V_r = 2.19$. Also, this maximum is found at a higher gate opening than the maximum of the closed gate ($0.74D$ against $0.48D$). This is the net effect of the altered gate design: a slightly different mass and additional discharge caused by the jet of relatively high velocity through the gate.

*2.5 Comparison with other experimental results*
The measurement data of the original rectangular gate resembles results from previous experimental research. It is important to realise that most other studies measure response in displacement instead of force. In the high stiffness at $V_r < 5$, which has been studied most frequently, the reduced velocity at which highest amplitude is found lies close to $V_r = 2.75$ found by Hardwick (1974) and $V_{r,nat} = 2.5$ found by Thang & Naudascher (1986). The reduced velocity based on the natural frequency $V_{r,nat}$ is somewhat lower than the $V_r$ used in this study. The gate openings of maximum amplitude were $a/D = 0.67$ and $a/D = 0.65$ for these two studies, slightly higher than what is found here. Billeter & Staubli (2000) report maximum force coefficients in z-direction at $V_r = 2.75$ and $V_r = 3.0$.

The experiments of Vrijer (1979) covered vibrations in the range $10 < V_r < 80$. Their model dimensions were smaller (gate width of 10 mm). A comparison based on the small overlap of less than ten data points suggests that the lower bound of vibrations in the low stiffness region in the present study is at a slightly lower $V_r$-value ($V_r = 9.0$-$9.5$) than in Vrijer (1979) where amplitudes start to increase for $V_r > 10$.

## *3. Numerical simulations*

*3.1 Model description*
The goal of the numerical simulations is to describe the physical features that are decisive for causing the difference in response between the two gate types. The numerical model is two-dimensional in the vertical direction (2DV) and has a two-way coupling between fluid and solid. Model dimensions are identical to the physical scale model. The flow-wise model length is 3.5 m. The static vertical equilibrium is achieved by a suspension support force equal to the difference between gravitational and buoyancy and mean lift forces. The flow velocity vector is defined as $\boldsymbol{u} = (u, v)$, the displacement vector of the solid body is written $\boldsymbol{u_s}$. The fluid flow is modelled by the incompressible Reynolds-Averaged Navier-Stokes (RANS) equations in combination with the standard $k$-$\varepsilon$ turbulence model. The solid domain includes deformation due to stresses. The governing equations are:

$$\rho \frac{\partial \overline{\boldsymbol{u}}}{\partial t} + \rho \overline{\boldsymbol{u}} \cdot \nabla \overline{\boldsymbol{u}} + \nabla \cdot \overline{(\rho \boldsymbol{u'} \otimes \boldsymbol{u'})} = -\nabla \bar{p} + \nabla \cdot \mu(\nabla \overline{\boldsymbol{u}} + (\nabla \overline{\boldsymbol{u}})^T) + \boldsymbol{F_g} \text{ and } \nabla \cdot \overline{\boldsymbol{u}} = 0 \text{ (fluid)}$$

$$\rho_s \frac{\partial^2 \boldsymbol{u_s}}{\partial t^2} - \nabla \cdot \boldsymbol{\sigma_s} = \boldsymbol{F_v} \text{ (solid)}$$

Where $\rho$ is the density of water, $p$ water pressure, $\mu$ is the viscosity of water, $\boldsymbol{F_g}$ the gravitational force and using Reynolds decomposition $\boldsymbol{u} = \boldsymbol{\bar{u}} + \boldsymbol{u'}$. Furthermore, $\rho_s$ is the solid's density, $\boldsymbol{\sigma_s}$ are the body stresses and $\boldsymbol{F_v}$ are the external forces on the body consisting of gravity, spring force, suspension force and the hydrodynamic forces. The upstream flow boundary condition is a block velocity profile and the downstream boundary is a hydrostatic pressure profile. The transient runs are preceded by stationary pre-runs with the gate held fixed in order to iteratively determine the inlet velocity at which the upstream pressure away from the gate is zero at the surface. Appendix B gives more details.

The Finite Element Method (FEM) is applied for spatial discretization. The Arbitrary Lagragian-Eulerian (ALE) method is used to make a moving mesh for the entire model domain (see e.g. Ferziger & Perić 2002). This means the grid adapts to the motion of the oscillating body without changing its connectivity. The free surface boundary is modelled as a rigid lid, which slightly bend up and down with the grid. A set of hyperelastic equations is solved to smooth the displacements of interior cells. The grid consists of unstructured triangular elements and inflated boundary refinements adjacent to the gate walls and flow bottom. A typical grid contained around 35,000 elements and 200,000 degrees of freedom. The cell size in the near-wall flow regions is dictated by the dimensionless wall distance by ensuring that $y^+_{max}$ = 11.06 at all walls where wall functions are applied (see Appendix B and COMSOL 2013). The online material contains a movie of the ALE moving mesh.

The FEM solver COMSOL Multiphysics v4.3a (COMSOL 2013) is used to solve the system of equations of seven dependent variables. The solution procedure is solved in a fully coupled way with the PARDISO direct solver. The implicit Backward Differentiation Formula (BDF) with adaptive time stepping is used, an extension of the backward Euler method for variable order. The simulations were done on a cluster, using 24 cores on a single node. Typically, three seconds of simulated time took around 12 hours of computing time.

Two cases from the experimental data set are selected for simulation; these are representative of strong vibrations in the low and high Vr region. See Table 1.

*Table 1. Selected cases for numerical simulations (values from experiment)*

|        | $Q$ (l/s) | $k$ (N/m) | $h_1$ (m) | $\Delta h$ (m) | $a/D$ (-) | $Vr$ (-) | $Cs$ (-) | $f_z$ (Hz) | $\hat{F}_z$ (N) |
|--------|-----------|-----------|-----------|----------------|-----------|----------|----------|------------|-----------------|
| **case 1** | 23 | 82502 | 0.398 | 0.088 | 1.00 | 2.6 | 5.2 | 10.1 | 25.6 |
| **case 2** | 40 | 19298 | 0.605 | 0.285 | 0.86 | 10.2 | 5.5 | 4.66 | 63.0 |

*3.2 Model validation results*
We are interested in the behaviour of both gates for conditions as close as possible to those found in the experiment. The followed approach was to assign applicable physical parameters from the experiment to the numerical model. Firstly, the added mass coefficient $m_w$ in still water is derived from a zero discharge model and compared to experimental values. Secondly, iterative validation runs are made in a model *with* discharge to achieve the settings necessary for attaining the response frequency as found in the experiment. These two preparatory modelling steps should be seen as efforts towards model validation (comparing the modelled value of a universal physical parameter with the experimental value) and calibration (adjusting the numerical model for the specific modelling task), respectively.

*3.2.1 Added mass validation*
The kinetic energy approach is followed to find $m_w$ in still water (see section 2.2). Because flow velocities are low, the laminar incompressible flow equations are solved. The oscillating gate is simulated by a moving wall with a prescribed vertical periodic velocity with a frequency of 4.7 Hz and amplitude of 0.15 m/s. The rigid lid assumption is justified for this situation because the wave radiation effect is small: for $h$ = 0.40m, we have $\omega^2 h/g = 36 \gg 10$, the usual criterion for wave radiation (Kolkman 1976).

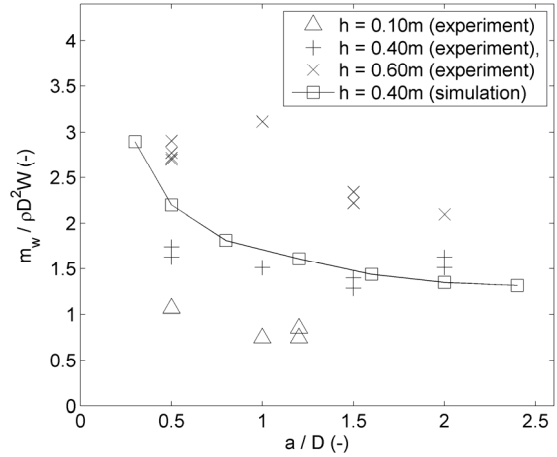

*Figure 4. Added mass coefficient in still water for various openings* a *and water depths* h.

In Figure 4 experimental values of non-dimensional added mass in still water are plotted for three water depths, together with simulation results for $h$ = 0.40 m. The added mass simulations show that the wall proximity effect at low $a/D$ is somewhat more pronounced than in the available experimental values. For vibrations at high gate openings, the presence of the bottom disappears. The simulations show a gradient reduction in correspondence with this, while experimental data at higher $a/D$ would be needed to cover this stabilisation. Experimental data by Nguyen (1982) in Naudascher & Rockwell (1994) show stabilisation of the added mass for $a/D$ > 2–3. The dependence on water depth that is shown by the experimental values is not sufficiently captured by the model: for other water depths, the results are too close to the plotted relation for $h$ = 0.40 m. The influence of domain width was neutralised by ignoring very small flow velocities in the computation. In summary, for the flow cases that are considered in this study the physical added mass is reasonably well approximated.

*3.2.2 Artificial added coefficients*
In the numerical model, the gate mass is defined by assigning a solid density $\rho_s$, and the suspension stiffness $k$ is defined in a spring suspension support. The added rigidity $k_w$ due to buoyancy is in all studied cases negligibly small, i.e. $k_w$ << $k$. Despite the fact that for oscillating bodies in flowing water $m_w$ and $k_w$ are in general not equal to their still water values (nor constants for that matter), it is reasoned that mismatches of simulated response frequences with experimental values, must be traced back to artificial added coefficients. Such misrepresentations were indeed found. It was checked that the natural frequency of the isolated gate in vacuum calculated by the same FEM model exactly matched the analytical value.

A possible cause of numerical complications in FSI is the so-called 'artificial added mass' effect. This has been investigated and partially described and explained for sequentially staggered schemes (Förster et al. 2007), but no literature was found for FSI with fully coupled schemes in FEM for relatively stiff solids. Similar to Jamal & Dalton (2004), we define the mass factor $m^*$ to be the ratio of solid density over fluid density. The term "mass ratio" is also used but can be confused with the dimensionless mass as appearing in Thang and Naudascher (1986) and in Figure 4, for example. The sensitivity of the simulated response frequency $f_{z,num}$ to the mass factor $m^*$ is plotted in Figure 5.

Using the true mass factor of the experiment, which is less than one, gives crude underestimates of the measured response frequency $f_{z,exp}$ for both investigated cases. The mass factor is increased to $m^*$ = 1.1 simply by using a lower gate height in the simulation than in the experiment. When $m^*$ is increased further by means of increasing solid density $\rho_s$, while at the same time adjusting $k$ proportionally such that the dry natural frequency $f_0$ remained constant, the numerical model gives better estimates of $f_{z,exp}$. Grid refinements did not noticeably influence the trends.

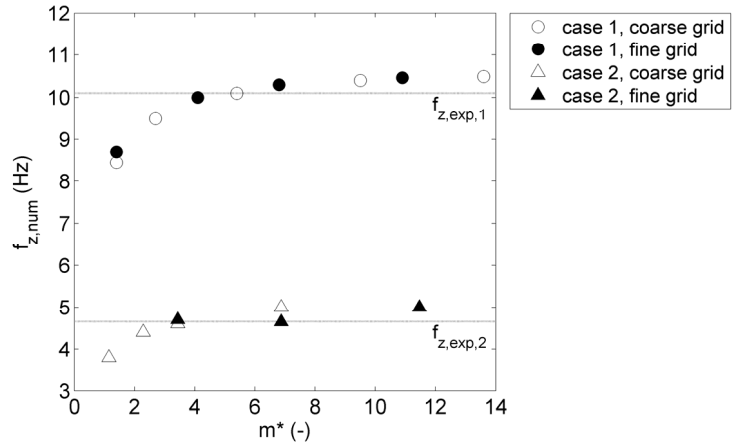

*Figure 5. The artificial added mass effect for the three cases of Table 1. All data points for each case have the same natural frequency. The dotted horizontal lines show the experimental response frequencies $f_{exp}$.*

Increasing $m^*$ at fixed $f_0$ does not yield an asymptotic approach to the experimental values, however. In addition, the achieved displacement amplitudes of the gate vibration are inconveniently small for $m^* \geq 10$ (order 0.1 mm and smaller). As a tradeoff it is decided to take $m^* = 3$ and adapt $k$ accordingly so that natural frequency and response frequency are well represented by the numerical model for both cases and displacements are in order of millimeters. It is noted that this measure distorts the absolute values of the modelled displacements.

*3.3 Results of calibrated numerical model*
Next, cases 1 and 2 were simulated with the calibrated numerical model. In each transient run, only three to four seconds were simulated. The numerical perturbations related to initialisation of the run were enough to kick-start a vibration – then the simulated vibration either quickly grew in amplitude or damped out. After a substantial number of test runs, no active response has been found for case 1. This is presumably due to an underestimation of the dynamics of the impinging shear layer, see section 4.1. For case 2 growing amplitudes were found at a slightly larger gate opening than in the experiment (46 mm in the simulation versus 43 mm in the experiment). The plots in this section are simulations of case 2, for the original and the ventilated gate. Movies of simulation case 2 in the online material show pressure, velocity and turbulent kinetic energy fields for the original gate and the ventilated gate type.

Figure 6 shows the simulated displacement time signals normalized to the gate width *D*. After an initially similar disturbance, the gate with holes follows a damped vibration and the original gate is amplified. The frequency of the ventilated gate is a little lower than the closed gate.

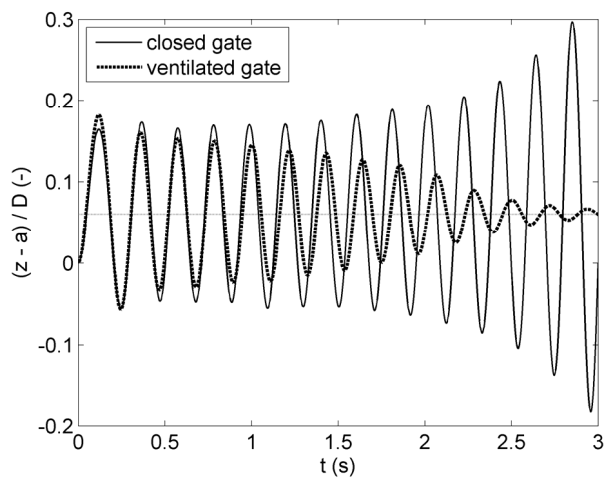

*Figure 6. Simulated vertical displacements of original, closed gate (thin continuous line) and gate with ventilation holes added (thick dashed line). Thin dotted horizontal line indicates the steady displacement of 3.0 mm experienced by both gates. The displacement is plotted relative to the original gate position z = a and normalized to the gate width D.*

Additional signals of the numerical simulations are plotted in Appendix C, giving lift force, displacement and acceleration in time. For one period starting at $t$ = 1.35 s, a number of plots is made in Figures 8-13 for further comparison of the modelled response of both gate types: left plots are the original gate and right plots the modified gate. Figure 7 gives the location of the output lines of pressure, velocity and turbulent kinetic energy profiles.

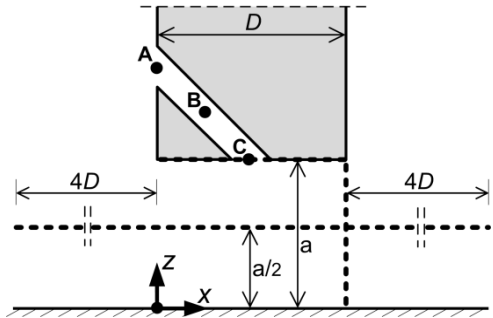

*Figure 7. Location of output profiles (dashed lines) and probe points used in Figures 8-13.*

Figures 8-13 not only show the effect of adding holes, but also allow comparison of both gate types with the situation where the gate is fixed. The variation of flow parameters is much smaller in the fixed gate scenario and therefore these are depicted as single thick dashed lines, which are time-averaged values. In each plot of Figures 8-12, the nine plotted lines correspond to different moments in one full sine period, according to $i\pi/4$ with $0 \leq i \leq 8$; these are indicated in the plots of the original gate. The fixed gate values in all cases lie between the extreme values of the moving gate scenario. The figures 8-13 show that the oscillating gate with holes experiences significantly lower periodic variation in streamwise pressure, velocity and turbulent kinetic energy (TKE) profiles at the trailing edge, and bottom pressure, than the gate without holes. Obviously this is associated with the difference in displacement amplitude at the time of output, but nevertheless an insightful comparison can be made.

The streamwise pressure plot (Figure 8) shows that, at a height of half the gate opening, the original gate experiences much larger temporal variation and higher maximum streamwise pressure gradients directly under the gate ($0 < x/D < 1$) than the ventilated gate. The latter shows low $|dp/dx|$-values for $x/D < 2$. The $u$-velocity profiles (Figure 9) clearly show the influence of the extra stream through the hole. The right plot contains a local flow maximum around $z = 0.8a$, which is absent in the left plot. Moreover, for the ventilated gate the area of negative streamwise velocity is very limited: $0.85 < z/a < 1$ versus $0.7 < z/a < 1$ for the original gate.

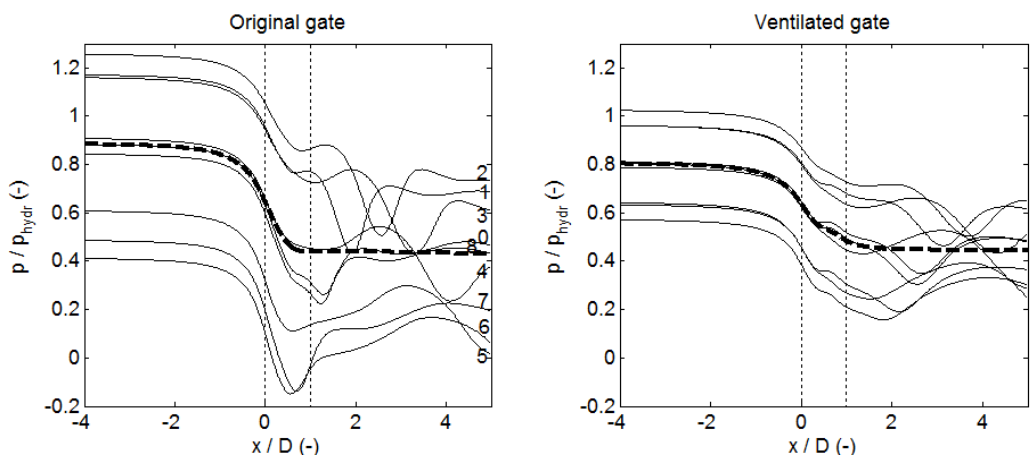

*Figure 8. Simulated streamwise pressure variation at z = a/2 for moving gate without (left) and with ventilation holes (right); $p_{hydr}$ is the hydrostatic pressure at z = a/2. Thick dashed lines show the situation where the gate is held fixed. Vertical thin dashed lines indicate location of the hole through the gate. The nine lines correspond to different moments in one full period, according to $i\pi/4$ with $0 \leq i \leq 8$.*

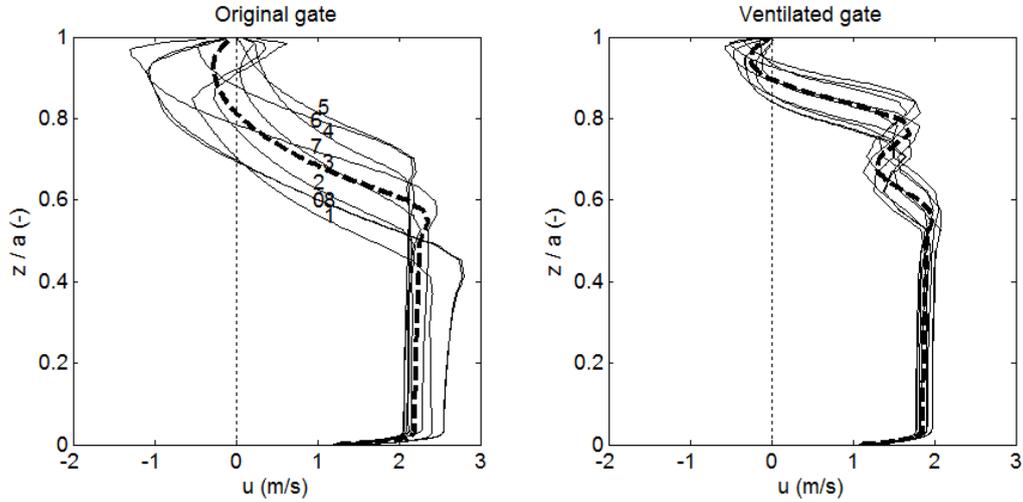

*Figure 9. Simulated vertical profile of horizontal velocity component* u *at downstream gate edge. (from bottom to gate). Thick dashed lines indicate the fixed gate scenario.*

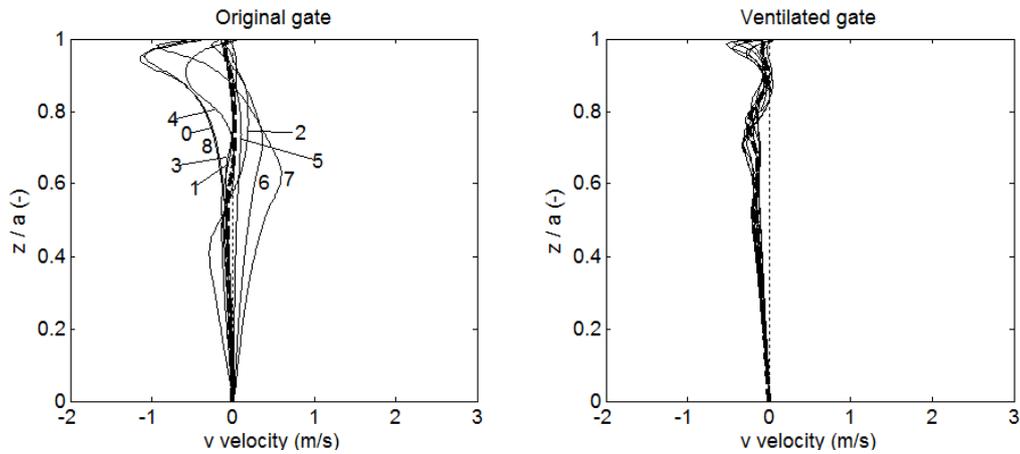

*Figure 10. Simulated vertical profile of vertical velocity component* v *at downstream gate edge. (from bottom to gate). Thick dashed lines indicate fixed gate scenario.*

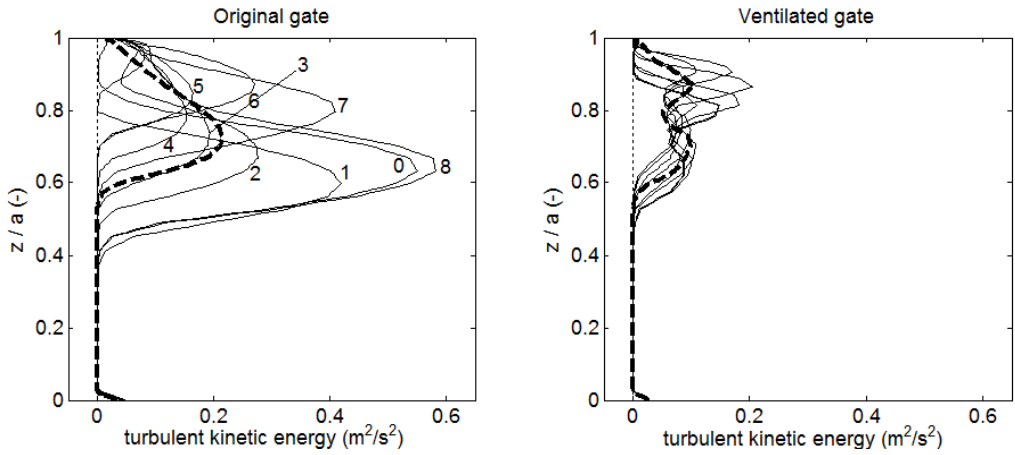

*Figure 11. Turbulent kinetic energy in vertical profile at the trailing edge of the gate.*

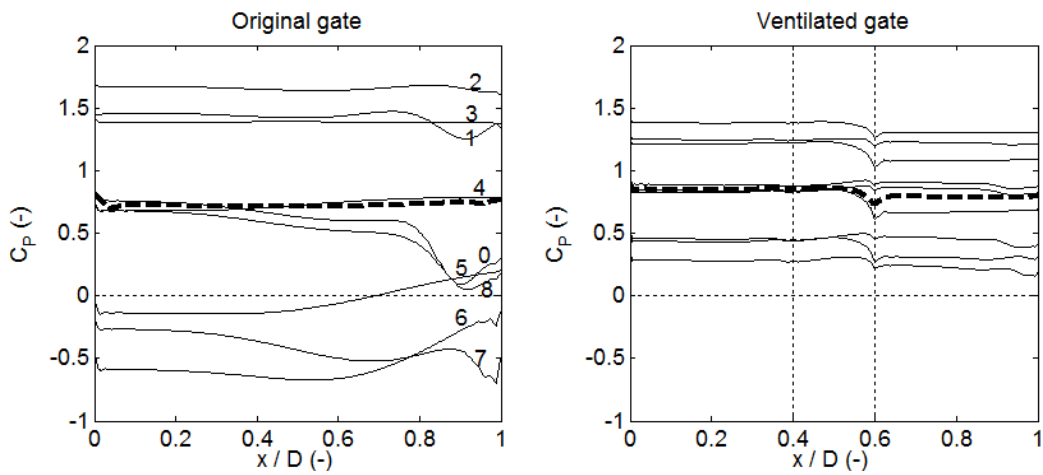

*Figure 12. Spatial variation of hydrodynamic pressure on the gate boundary for both gate types. The vertical dashed lines show the location of the hole.*

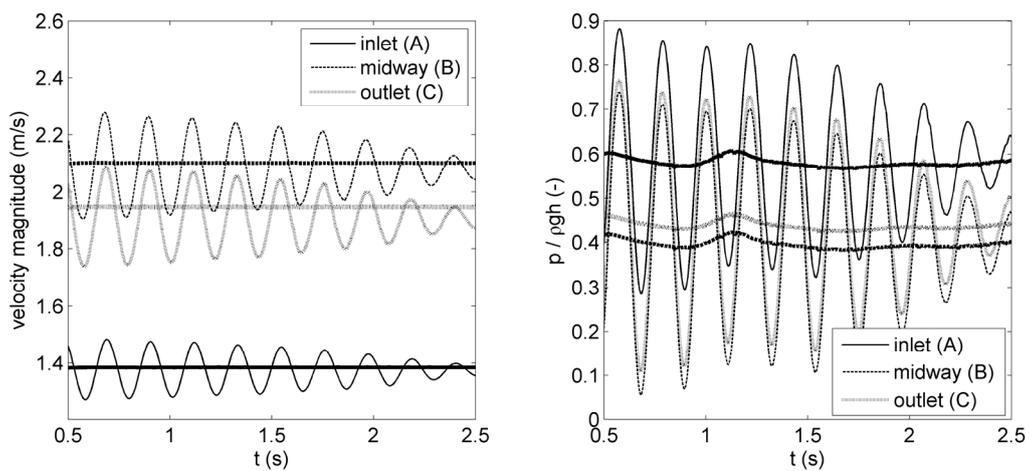

*Figure 13. Velocity magnitude and pressure relative to hydrostatic pressure at z = a at three points inside the ventilation hole. Points A-C refer to the locations in Figure 7. The bold lines are the values when the gates are held fixed.*

The effect of the downward jet though the hole on *v*-velocity (Figure 10) is that the flow direction is now almost completely downward during the whole oscillation period. The TKE plots in Figure 11 show that the TKE maximum is greatly reduced and located closer to the gate. In the ventilated case, two local maxima of turbulent kinetic energy are discerned, most clearly visible for the fixed gate. These correspond to the two shear layers resulting from flow separation from the downstream edge of the hole outlet (maximum around z = 0.85$a$) and from the leading edge of the gate bottom (maximum around $z$ = 0.7$a$).

The boundary pressures in Figure 12 indicate that the reduced temporal fluctuation for the new gate results in positive pressures during the whole cycle, whereas the boundary pressure on the original gate is negative in the last quarter of the cycle (between lowest position and equilibrium position). The spatial pressure variations towards the trailing edge seem to signify unstable reattachment for the original gate. This is virtually absent for the gate with holes. There are only small pressure drops visible near the outflow of the ventilation hole.

Figure 13 plots the velocity magnitude and pressure of the flow inside the hole. The periodic variation of these flow properties is fully related to the gate's motion – the frequency equals that of the displacement. The flow accelerates in the upper part of the hole. Closer inspection learns that around the inlet locally high pressure gradients, velocities and TKE-values occur. This can be improved in future studies by streamlining the intake geometry.

Appendix D gives more simulation results of both gate types for different moments in time for one oscillation period of case 2. Additionally, plots are shown for case 1 for the original gate type and for the fixed gate scenario for both gate types. The plots of pressure contour lines and flow velocity vectors serve to point out qualitative differences between the two gate types. The maximum velocities are indicated. It can be seen that the flow through the hole works to disturb the separated shear layer and to keep the associated instabilities away from the gate. The reversion of streamwise flow velocity at the bottom boundary and instantaneous spatial pressure gradients near the trailing edge appear for the gate without holes in plots (3) and (4). For the gate with holes these two features less significant. As already mentioned, the ventilated gate is characterised by the existence of two separation points and two shear layers. The jet through the hole persists in its downward flow during the whole oscillation period. This causes a relief in pressure built up, see plots (1) and (5).

The simulation of case 1 shows negative *u*-velocity at the gate boundary for the whole period. The fact that even at very small gate motion pressure contours are colliding with the boundary, see plots (1) and (4), can be seen as evidence for unstable reattachment of the shear layer. Comparing the situations of cases 1 and 2 directly is difficult because of the much lower simulated displacement amplitude of case 1.

## *4. Discussion*

*4.1 Vibration mechanisms*
As explained in the introduction, past research largely agrees on the existence and working of two excitation mechanisms for cross-flow gate vibrations: ILEV and MIE (noise excitation or Extraneously Induced Excitation (EIE) is not considered here). The instability-induced or vortex-excited type is the most likely mechanism for the vibrations found around $Vr \approx 2.5$, while the movement-induced type is probably the dominant mechanism for the vibrations at $Vr \approx 9.5$.

The limited numerical modelling result of case 1 does not exclude the possibility that unstable reattachment of the separated shear layer (i.e. ILEV) could be the main cause of excitation in the simulation. The separation point is fixed at the leading edge – irrespective of discharge and gate position – and the boundary pressure fluctuates near the trailing edge, both in space and time. However, given the fact that the strongest vibrations in the experiment occurred for conditions similar to case 1 (viz. at $Vr \approx$ 2.5), it is remarkable that the simulation of case 1 vibrations damped out. Runs with adapted settings in the neighbourhoud of case 1 (higher or lower gate opening, higher or lower discharge) were attempted, but did not yield different results. A transient simulation of case 1 conditions with the gate held fixed showed only small flow instabilities in the shear layer, making it impossible to determine the Strouhal number and identify small-scale vortices.

Simulation case 2 displays the main feature of excitation by movement (MIE) or self-excitation, because not only is a motion required to initiate the vibration process, the fluctuations in the boundary excitation force do not appear at all in the transient simulation when the gate position is fixed. On the other hand, the phase shift $\varphi$ between excitation force and displacement is close to zero and the mean lift is only negative for a quarter of the period. Therefore, there is no direct evidence of galloping-type MIE excitation (Billeter & Staubli 2000). The mechanism of energy transfer to the motion thus stems not just from the flow itself but possibly from the fact that water level difference and discharge are constant in the numerical model.

Kolkman & Vrijer (1987) reported on a mechanism based on streamwise flow inertia caused by the discharge not instantly adapting to the new gate position during low frequency vibrations at small gate openings. Because the discharge is never choked in the numerical model, this mechanism is by definition not simulated. However, the role of this mechanism cannot easily be put aside, judging from the prominent streamwise pressure irregularities in the simulations.

The weakening effect of the added holes in the case of vibration due to ILEV consists of a disruption of the separated shear layer and vortices shed from the leading edge. The downward flow through the bottom removes instabilities from the gate bottom and hence reduces the possibility and impact of unstable reattachment. In the case of self-excited vibration or MIE, the most likely working of the flow through the holes is that it alleviates the fluctuation of the streamwise pressure and the fluctuation of the boundary pressure (thus reducing or removing negative lift).

The jet that flows out the middle of the bottom boundary not only removes local disturbances from the vicinity of the gate, but also has the effect of splitting the flat bottom surface into two parts as if there were two thin gates. This is favourable since thinner profiles are less prone to vibrations (or at least in a smaller region of gate openings). An alternative explanation of the response reduction could be found in the fact that the presence of the hole simply decreases the surface area for the lift pressure to work on. Simulations of a closed rectangular gate with a bottom surface area equivalent to that of the gate with a hole showed that the diminished dynamic force amplitude cannot be explained by a smaller area. The different response must therefore be caused by a change in the vibration mechanism(s).

*4.2 Evaluation of numerical modelling*
A number of aspects of the numerical model are briefly discussed. The rigid lid approach is a notable assumption that always requires some consideration. Wave radiation is a free-surface phenomenon that is not captured because of the rigid lid. It is a form of hydraulic damping that involves vibration energy to be transformed irreversibly into free-surface waves. Not modelling this is not a severe omission, see also section 3.2. In the experiment wave radiation was only observed once at the upstream water surface in the low stiffness vibration region. Another free-surface effect that is completely neglected by using the rigid lid approximation is the coupling of flow-induced undulations of the downstream surface with the gate oscillation at increasing Froude number (see Naudascher & Rockwell 1994). However, these undulations only occur at much lower submergence levels than investigated in this study. A consequence of the fixed free surface boundary in connection with the constant discharge is furthermore that global pressure oscillations are exaggerated. A lowering of the gate results in an immediate pressure increase in the upstream region, for instance. It is not easily found to what extent this affects the emergence of gate instabilities in the simulations; there is presumably a link with the representation of the MIE vibrations. The complication of this numerical modelling aspect lies in the fact that methods that do simulate the free surface (such as Volume Of Fluid or Phase Field) are usually not suitable for including both moving objects and turbulent flow, or they are computationally impractically expensive.

In physical experiments, comparisons of measured gate displacement and excitation force cannot always distinguish between added mass and rigidity since both coefficients are part of terms in-phase with the displacement (e.g. Kolkman 1984). Similarly, in a numerical study it is not sufficiently clear whether a deviating response frequency should be attributed to artificial mass or rigidity effects. Either way, the distorting influence of artificial added coefficients must be elucidated before the full working of physical mechanisms can be uncovered by numerical modelling.

Artificial damping or numerical diffusion can be traced back to the use of an implicit time scheme and 'consistent stabilisers' (COMSOL 2013). The effect of this is inherent to the followed numerical approach; without it the simulations fail. Naturally, too much damping results in vibrations not being induced in the simulations where they do occur in real life. In particular, negative damping plays a key role in self-excited

vibrations – if negative damping is neutralised by artificial factors, MIE-type vibrations will be inaccurately represented.

Finally, the obvious limitation of the RANS approach with a turbulence model is that not all turbulence scales are simulated. The fact that vortex shedding and velocity and pressure fluctuations at small-length scales connected to the shear layer are parametrised too coarsely could be the reason why the ILEV-dominated vibrations of case 1 were not reproduced. Applying a different turbulence model will probably not improve this; Large Eddy Simulation (LES) could be the only way forward.

## *5. Conclusions and future work*

The aim of this study was to experimentally and numerically test a new hydraulic gate design for reducing flow-induced cross-flow vibrations. The addition of ventilation holes to a rectangular flat-bottom gate allowed a controlled leakage flow through the bottom of the gate, which produced a less severe vibration response as compared to an unaltered reference gate.

The experimental data set fully covers the transitions between conditions with and without significant flow-induced vibrations in the reduced velocity region $2 < Vr < 3.5$. Two distinct vibration regimes are recognized with maximum response force amplitudes at $Vr = 2.54$ and $Vr = 10.16$ for gate openings $a/D = 0.48$ and $a/D = 0.86$, respectively. The results show that the gate with perforated bottom profile significantly reduces cross-flow vibrations in the region $2 < Vr < 3.5$. Although not exhaustively covered, the measurements give reason to believe that the same holds for the higher region at $Vr > 8.5$.

The obtained data set was used to evaluate and improve the performance of a numerical model. Time-dependent FEM simulations on a moving grid were performed to solve the RANS equations for the flow and the gate displacement of the mass-spring system. An initial model validation step for the added mass in still water showed reasonable estimates compared to the measurements and reproduced the wall proximity effect. Subsequently, as a form of calibration, the sensitivity of the response frequency in flowing water to variation of the mass factor was used to select an appropriate solid density and spring stiffness while keeping the natural frequency in air equal to experimental values. Then, two cases of strong vibrations were simulated from the low and high $Vr$ regions. For the high $Vr$ case with response frequency 4.7 Hz, vibrations with growing amplitude were reproduced for the rectangular gate while the modified gate with a hole showed decreasing amplitudes in the same conditions. This matches results from the physical experiment.

Even though the numerical model does not capture all features relevant for a complete representation of the excitation mechanisms, the results nevertheless give valuable information about the working of the new gate design. The downward jet through the hole removes flow instabilities of the leading edge shear layer from the bottom boundary and smoothens local pressure gradients. The combined results of measurements and numerical computations lead to the conclusion that the application of leakage flow is a promising new way to reduce the effect of flow-induced gate vibrations at small gate openings.

The quantification of the fluid-structure interaction of gates is a challenging task. The approach followed in this study in which a physical experiment is complemented with physics-based numerical modelling is considered to be suitable for gaining understanding of gate design performance and excitation mechanisms. Future research should examine the new gate design in a wider range of conditions, most notably $Vr > 9.5$. It should also improve flow angle, placement and design of the holes and see how to fine-tune the leakage flow for optimisation of the reduction effect. Moreover, practical issues should be identified and addressed before real-life application is possible. For numerical modelling it is recommended to develop techniques that are able to combine FSI with simulations of the free surface and detailed turbulent flow.

Finally, the introduction of informatics in the field of hydraulic engineering has led to the application of data-driven modelling (Solomatine & Ostfeld 2008). Obviously, this approach does not provide a direct way towards a more profound understanding of physical mechanisms, but instead it can be very powerful in gate control systems. Erdbrink et al. (2012b) presented the concept of a data-driven system to prevent flow-induced vibrations based on data from sensors on the gates. A similar proposal has been reported by Han et al. (2011). The signals of the experimental data-set of this study are suitable for training a data-driven model to recognise transition characteristics when the gate approaches a region of high amplitude vibrations.

## *Acknowledgements*

The experiment was carried out at Deltares as part of project 1202229-004. We thank Tom Jongeling for his guidance in designing the measurement set-up and the supporting personnel at Deltares. This work is supported by the Leading Scientist Program of the Russian Federation, contract 11.G34.31.0019.

## Appendix A    Experimental set-up: flume

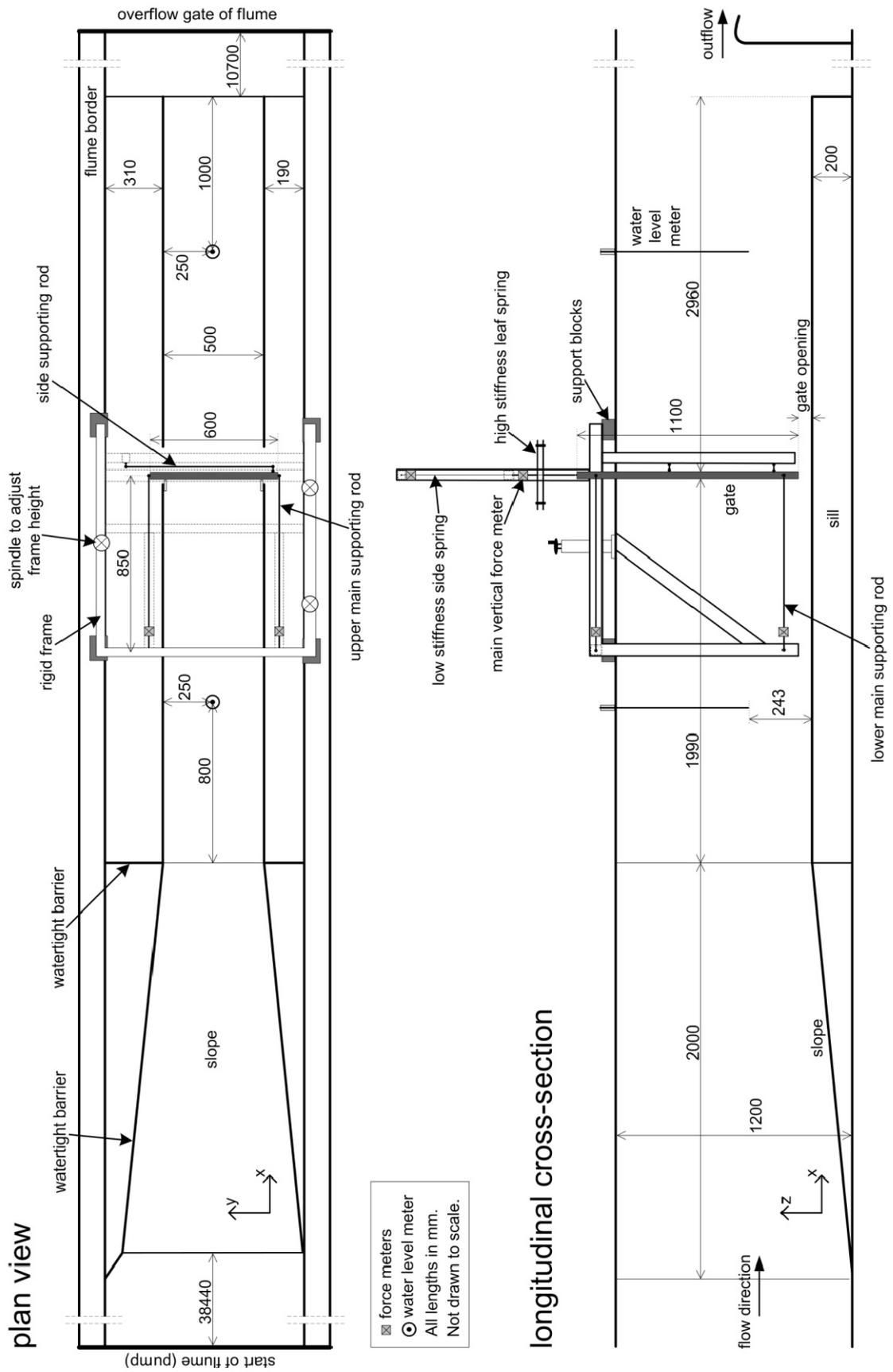

*Figure 14. Sketched drawings of experimental set-up in flume.*

*Table 2. Measurement conditions (flume settings)*

|  | min | max | unit |
|---|---|---|---|
| pump discharge $Q_{pump}$ | 13.0 | 50.0 | l/s |
| gate opening $a$ | 22.5 | 100 | mm |
| upstream water depth $h_1$ | 0.297 | 0.656 | m |
| downstream water depth $h_3$ | 0.120 | 0.352 | m |
| head difference $\Delta h$ | 0.063 | 0.355 | m |

Parameter ranges given in Table 2 are based on the combined data set for both gates, consisting of 145 measurements for the original gate plus 85 measurements for the modified gate. The achieved gate discharge (and computed discharge coefficient) are slightly influenced by a sideways leakage at high upstream water depths, through the rubber side seals. An observed variation in the pump discharge of +/- 0.5 l/s was of little influence since the frequency of this variation was very small compared to signal recording length.

## *Appendix B      Numerical model equations*

**Fluid domain**

Time-dependent Reynolds-Averaged Navier-Stokes (RANS) equations for incompressible flow with $k$-$\varepsilon$ turbulence model are used:

$$\rho \frac{\partial \overline{\boldsymbol{u}}}{\partial t} + \rho \overline{\boldsymbol{u}} \cdot \nabla \overline{\boldsymbol{u}} + \nabla \cdot \overline{(\rho \boldsymbol{u}' \otimes \boldsymbol{u}')} = -\nabla \bar{p} + \nabla \cdot \mu(\nabla \overline{\boldsymbol{u}} + (\nabla \overline{\boldsymbol{u}})^T) + \boldsymbol{F}_g$$

$$\nabla \cdot \overline{\boldsymbol{u}} = 0$$

where $\rho$ is density of water, $\mu$ is the dynamic viscosity of water, $p$ is the water pressure, bars indicate time-averaged values and primes indicate variational variables

$$\overline{\boldsymbol{u}} = \begin{pmatrix} \bar{u} \\ \bar{v} \end{pmatrix}, \text{flow velocities}$$

$$\boldsymbol{F}_g = \begin{pmatrix} 0 \\ -\rho g \end{pmatrix}, \text{gravity (force per unit volume)}$$

The $k$-$\varepsilon$ turbulence model provides closure by solving a coupled pair of PDE's for turbulent kinetic energy $k$ and turbulent dissipation $\varepsilon$ (see COMSOL 2013).

*Boundary conditions*

At the inlet a block velocity: $\boldsymbol{u} = \begin{pmatrix} U_0 \\ 0 \end{pmatrix}$, with $U_0$ a chosen constant.

At the outlet a hydrostatic pressure profile is imposed: $p(y) = (h_{out} - y)\rho g$, where $h_{out}$ is the water depth at the outlet, equal to the height of the flow domain.

Bottom flow boundary: $\boldsymbol{u} = 0$, no slip with a wall function describing the near-wall velocity profile.

Boundary at water surface: $\boldsymbol{u} \cdot \boldsymbol{n} = 0$, free slip or 'rigid lid'.

*Initial conditions*

At $t = 0$, $p = 0$ for all $(x, y)$, and $\boldsymbol{u} = \begin{pmatrix} U_0 \cdot step(t) \\ 0 \end{pmatrix}$, with step($t$) a smooth S-curve increasing from 0 to 1 for 0 < $t$ < 0.5 s.

*Wall function condition*

The wall lift-off (dimensionless wall distance) is defined as

$$y^+ = \frac{\rho u_\tau y}{\mu} \text{ with shear or friction velocity } u_\tau = C_\mu^{1/4}\sqrt{k}$$

The condition for full resolution is $y^+ = 11.06.$, the distance at which the viscous sub-layer and the logarithmic layer meet (see e.g. COMSOL 2013).

**Solid domain**

Defining $\boldsymbol{u}_s = (u_{s,x}; u_{s,y})$ as the displacement vector of the solid body and $\boldsymbol{\sigma}_s$ as the stresses it experiences (a tensor), we have

$$\rho_s \frac{\partial^2 \boldsymbol{u}_s}{\partial t^2} - \nabla \cdot \boldsymbol{\sigma}_s = \boldsymbol{F}_v$$

The applied external forces $\boldsymbol{F}_v$ acting on the gate body are

$$\boldsymbol{F}_v = \boldsymbol{F}_g + \boldsymbol{F}_{spring} + \boldsymbol{F}_{water} + \boldsymbol{F}_{suspension} = \begin{pmatrix} 0 \\ -\rho_s g \end{pmatrix} + \begin{pmatrix} 0 \\ -k_{spring}u_{s,y} - c_{spring}\dot{u}_{s,y} \end{pmatrix} + \boldsymbol{F}_{water} + \boldsymbol{F}_{suspension}$$

Where $F_{water}$ respresents the dynamic load of the water flow on the submerged part of the solid, see below. The spring force depends on chosen constants for stiffness, $k_{spring}$, and damping, $c_{spring}$. The last force represents the steady suspension force in the structural element that connects the gate object with the bigger (fixed) structure.

The constitutive relation (stress-strain relation) can be written as $\sigma_s = \sigma_0 + C:(\varepsilon - \varepsilon_0)$ with sub index 0 indicating initial values; $C$ is the elasticity, a function of the material properties (Young's modulus and Poisson ratio). The strain tensor is given by

$$\boldsymbol{\varepsilon}_s = \frac{1}{2}(\nabla \boldsymbol{u}_s + \nabla \boldsymbol{u}_s^T), \text{ which is equivalent to } \varepsilon_{xx} = \frac{\partial u}{\partial x}, \; \varepsilon_{yy} = \frac{\partial v}{\partial y} \text{ and } \varepsilon_{xy} = \varepsilon_{xy} = \frac{1}{2}\left(\frac{\partial u}{\partial y} + \frac{\partial v}{\partial x}\right).$$

*Boundary conditions*
Displacement is fixed in horizontal direction: $u_{s,x} = 0$

*Initial conditions*
At $t = 0$,
$$\boldsymbol{u}_s = 0 \text{ and } \frac{\partial \boldsymbol{u}_s}{\partial t} = 0$$

**Fluid-solid interface**
*Boundary condition*
The load that the water flow forces exert on the boundaries of the solid body, $F_{water}$, is defined as a force per unit area (in N/m²):

$$\boldsymbol{F}_{water} = \boldsymbol{T} \cdot \boldsymbol{n}$$

where $\boldsymbol{n}$ is the normal vector on the solid boundary and $\boldsymbol{T}$ is the 2x2 tensor containing all stresses of the flow. The normal stresses, on the diagonal of $\boldsymbol{T}$, represent the pressure. The remaining elements of $\boldsymbol{T}$ represent the viscous shear stresses – each of these elements consist of a stress contribution from the turbulent viscosity ($\mu_T$) and the laminar viscosity ($\mu$).

*Initial conditions*
At $t = 0$, $\boldsymbol{F}_{water} = 0$, and subsequently $\boldsymbol{F}_{water} = \boldsymbol{T} \cdot \boldsymbol{n} \cdot step(t)$, with the same S-function increasing from 0 to 1 for $0 < t \leq 0.5$ s.

**Moving mesh**
The whole domain (i.e. fluid and solid together) is subject to free mesh deformation. The boundary conditions for the mesh displacement vector $\boldsymbol{u}_m$ are

$$u_{m,x} = 0 \text{ on the fluid surface boundaries}$$
$$\boldsymbol{u}_m = 0 \text{ on all remaining boundaries of the fluid domain}$$
$$\boldsymbol{u}_m = \boldsymbol{u}_s \text{ on the solid boundaries}$$

For the inner area, the mesh deformation is smoothed using hyper elastic smoothing, which is similar in form to the description of neo-Hookean materials (see COMSOL 2013). This approach seeks for a minimum of 'mesh deformation energy' expressed as

$$W = \int_\Omega \frac{\mu}{2}(I_1 - 3) + \frac{\kappa}{2}(J - 1)^2 dV$$

Where $\mu$ and $\kappa$ are the equivalent constants representing shear and bulk moduli. Furthermore, $J$ and $I_1$ are invariants:
$$J = det(\nabla_X x) \text{ and } I_1 = J^{-2/3} tr((\nabla_X x)^T \nabla_X x)$$

Where $det(A)$ denotes the determinant of matrix $A$ and $tr(A)$ denotes the trace of matrix $A$.

# Appendix C     Numerical simulations: gate response in time

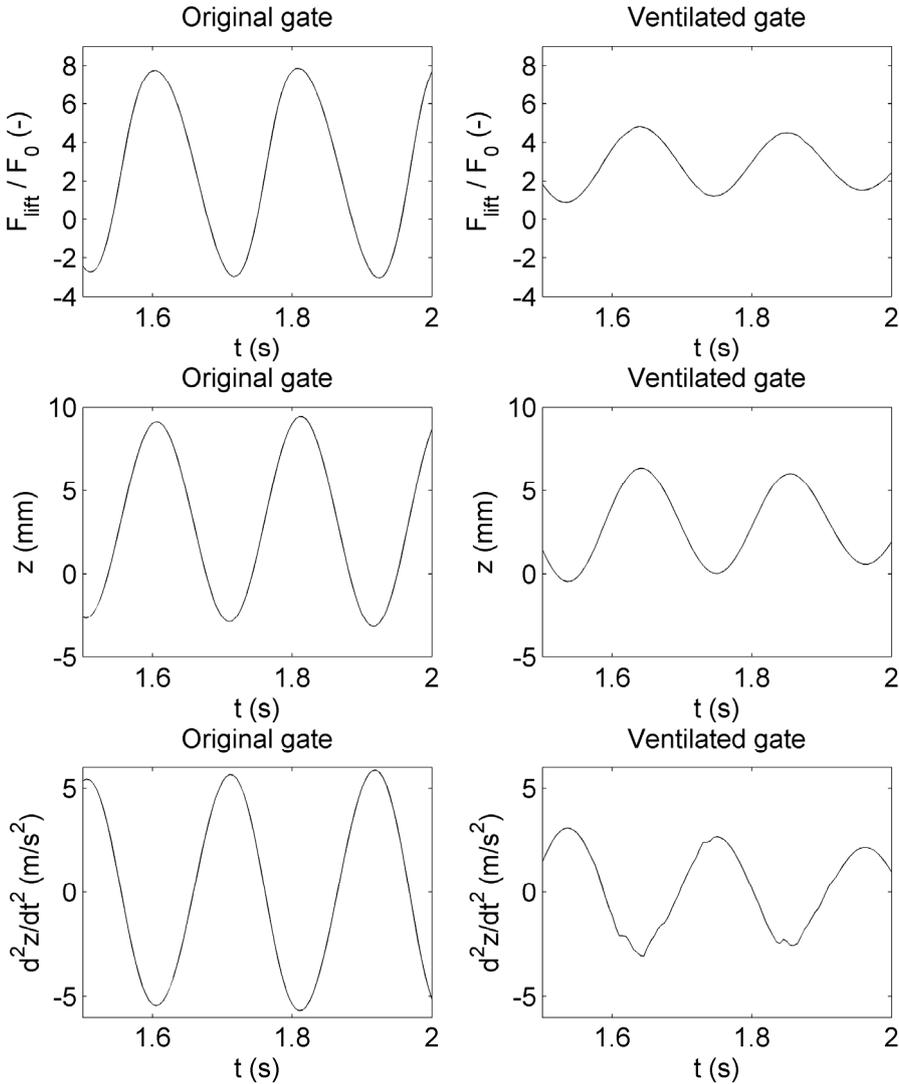

*Figure 15. Simulation case 2: time signal excerpts of dimensionless lift force and acceleration of the gate, for both gate types.*

# Appendix D  Numerical simulations: flow velocity and pressure

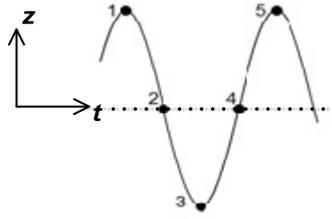

**Case 2: original gate, $f_z$ = 4.7 Hz**  |  **Case 2: ventilated gate, $f_z$ = 4.7 Hz**

(1)
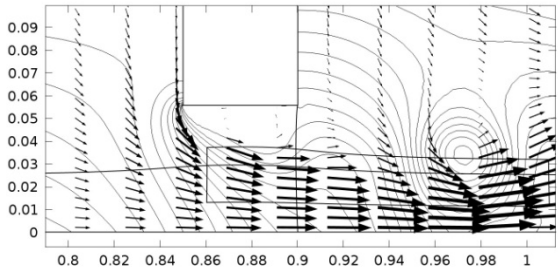 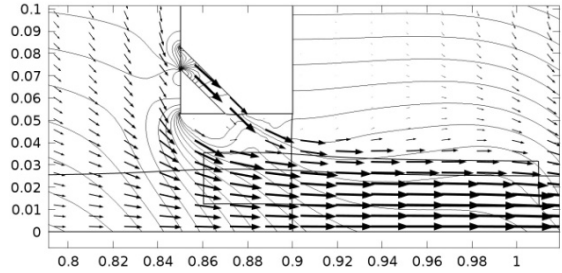

(2)
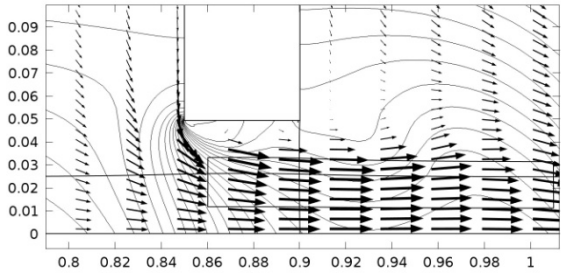 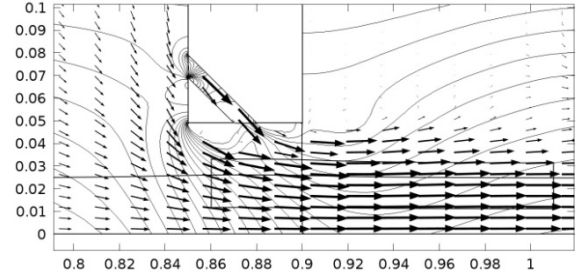

(3)
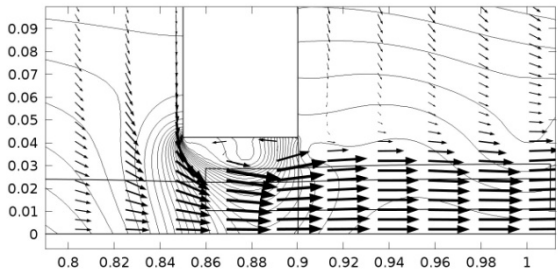 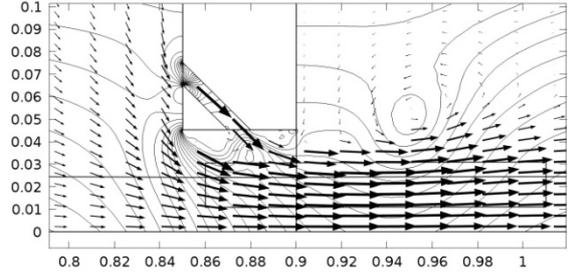

(4)
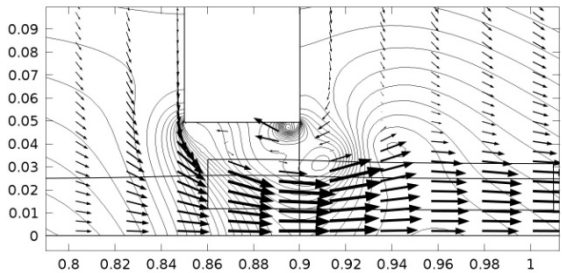 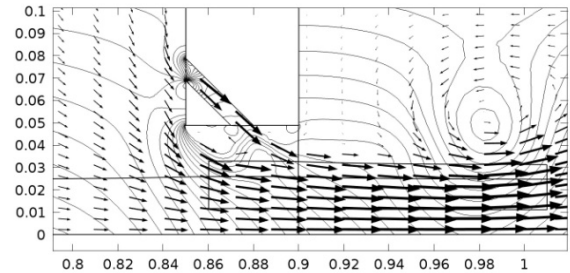

(5) 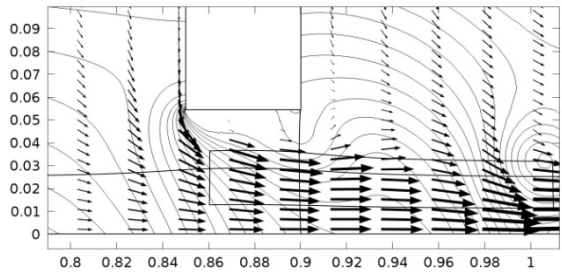 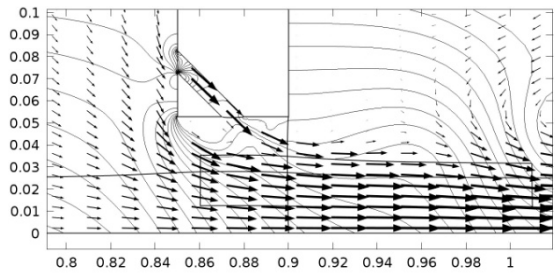

$U_{max, original, moving}$ = 3.38 m/s (highest global $U_{max}$) at (2) and 2.87 m/s (lowest global $U_{max}$) at (4).
$U_{max, ventilated, moving}$ = 3.22 m/s (highest $U_{max}$) at (2) and 2.42 m/s (lowest global $U_{max}$) at (4).

### Case1: original gate at $f_z$ = 10.1 Hz

(1). 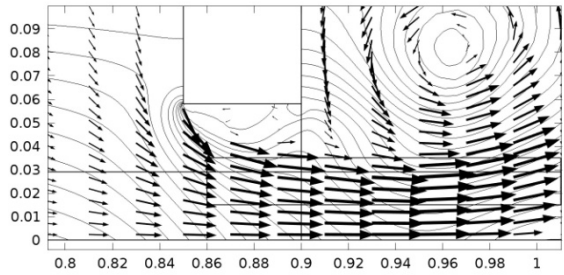    (2). 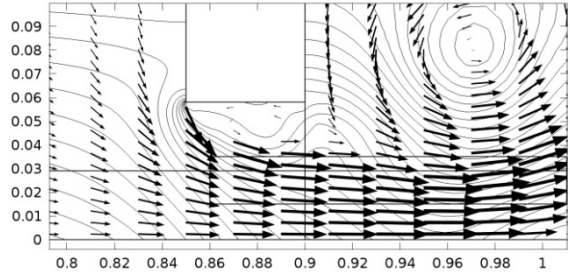

(3). 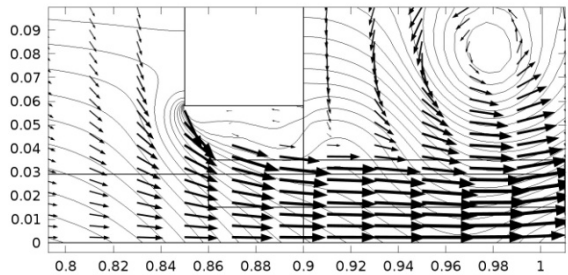    (4). 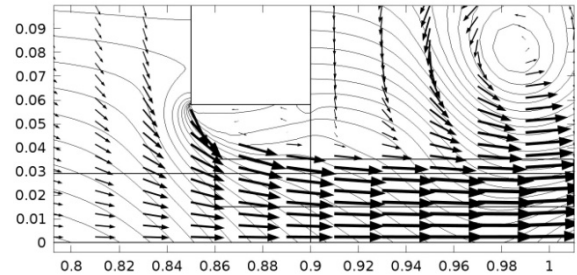

(5). 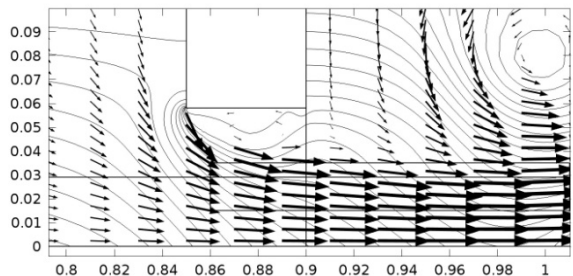

| **Original gate in fixed position*** | **Ventilated gate in fixed position*** |
|---|---|
| 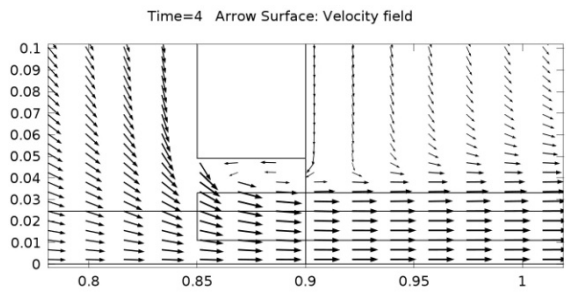 | 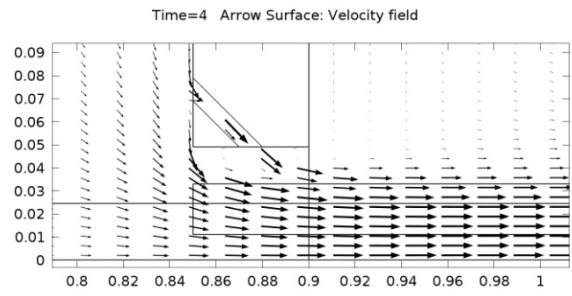 |

$U_{max, original, fixed}$ = 2.38 m/s
$U_{max, ventilated, fixed}$ = 2.36 m/s

* Flow conditions are equal to case 2.